\documentclass[a4paper,fleqn]{cas-dc}

\usepackage[square,sort&compress,numbers]{natbib}
\usepackage{tikz}
\usepackage{mathtools}
\usepackage{scalerel}
\usepackage{amsmath}
\usepackage{cuted}
\usepackage{caption}
\usepackage{subcaption}
\usetikzlibrary{arrows,arrows.meta,snakes,backgrounds,decorations.pathreplacing,decorations.markings}

\def\genbox#1#2#3#4#5#6{% #1=0/1, #2=color, #3=shape, #4=raise, #5=width, #6=width/2
    \leavevmode\raise#4bp\hbox to#5bp{\vrule height#5bp depth0bp width0bp
    \pdfliteral{q .5 w \csname #2COLOR\endcsname\space RG
                       \csname #3PDF\endcsname{#5}{#6} S Q
             \ifx1#1 q \csname #2COLOR\endcsname\space rg 
                       \csname #3PDF\endcsname{#5}{#6} f Q\fi}\hss}}
\def\circbox    #1#2{\genbox{#1}{#2}  {circ}     {0}   {5}    {2.5}}
\def\trianbox   #1#2{\genbox{#1}{#2}  {trian}    {0}   {5}    {2.5}}

\def\tsc#1{\csdef{#1}{\textsc{\lowercase{#1}}\xspace}}
\tsc{WGM}
\tsc{QE}
\tsc{EP}
\tsc{PMS}
\tsc{BEC}
\tsc{DE}
\begin{document}
\let\WriteBookmarks\relax
\def\floatpagepagefraction{1}
\def\textpagefraction{.001}

%%%%%%%%%%%%%%%%%%%%%%%%%%%%%%%%%%%%%%%%%%%%%%%%%%%%%%%%%%%%%%%%%%%
% Short title
\shorttitle{RSRG in CC model: improved statistics}

% Short author
\shortauthors{Syl Shaw and Rudolf A. Römer}

% Main title of the paper
\title[mode = title]{Real-space renormalisation approach to the Chalker-Coddington model revisited: improved statistics}                      
% Title footnote mark
% eg: \tnotemark[1]
%\tnotemark[1,2]

% Title footnote 1.
% eg: \tnotetext[1]{Title footnote text}
% \tnotetext[<tnote number>]{<tnote text>} 

\author[1]{Syl Shaw}[type=editor,orcid=0009-0007-9239-4092]

% Corresponding author indication
\cormark[1]

% Email id of the first author
\ead{syl.shaw@warwick.ac.uk}

%  Credit authorship
\credit{Conceptualisation, Investigation, Data curation, Formal Analysis, Software, Visualisation, Writing - original draft}

% Address/affiliation
\affiliation[1]{organization={Department of Physics},
    addressline={University of Warwick, Gibbet Hill Road}, 
    city={Coventry},
    % citysep={}, % Uncomment if no comma needed between city and postcode
    postcode={CV4 7AL}, 
    % state={},
    country={United Kingdom}}

% Second author

\author[1]{Rudolf A. Römer}[type=editor,orcid=0000-0002-8719-3313]
\ead{r.roemer@warwick.ac.uk}
\credit{Conceptualisation, Methodology, Writing - review \& editing, Supervision}

%%%%%%%%%%%%%%%%%%%%%%%%%%%%%%%%%%%%%%%%%%%%%%%%%%%%%%%%%%%%%%%%%%%
% Here goes the abstract
\begin{abstract}
The real-space renormalisation group method can be applied to the Chalker-Coddington model of the quantum Hall transition to provide a convenient numerical estimation of the localisation critical exponent, $\nu$. Previous such studies found $\nu\sim 2.39$ which falls considerably short of the current best estimates by transfer matrix ($\nu=2.593 \substack{+0.005\\-0.006}$) and exact-diagonalisation studies ($\nu=2.58(3)$). By increasing the amount of data $500$ fold we can now measure closer to the critical point and find an improved estimate $\nu=2.51\substack{+0.11\\-0.11}$. This deviates only $\sim 3\%$ from the previous two values and is already better than the $\sim 7\%$ accuracy of the classical small-cell renormalisation approach from which our method is adapted.
We also study a previously proposed mixing of the Chalker-Coddington model with a classical scattering model which is meant to provide a route to understanding why experimental estimates give a lower $\nu\sim 2.3$. Upon implementing this mixing into our RG unit, we find only further increases to the value of $\nu$.
\end{abstract}
%%%%%%%%%%%%%%%%%%%%%%%%%%%%%%%%%%%%%%%%%%%%%%%%%%%%%%%%%%%%%%%%%%%

% Use if graphical abstract is present
% \begin{graphicalabstract}
% \includegraphics{figs/grabs.pdf}
% \end{graphicalabstract}

% Research highlights
\begin{highlights}
\item We measure the critical exponent of the quantum Hall transition, $\nu$, ten times closer to the critical point by improving upon previous numerics with a $500$fold increase in the number of data points.
\item We find an increased value of $\nu=2.51\pm 0.11$.
\item Our new value is just $3.2\%$ different from the current best numerical prediction based on large-scale transfer matrix computations.
\end{highlights}

% Keywords
% Each keyword is separated by \sep

\maketitle

%%%%%%%%%%%%%%%%%%%%%%%%%%%%%%%%%%%%%%%%%%%%%%%%%%%%%%%%%%%%%%%%%%%
\section{Introduction}
%%%%%%%%%%%%%%%%%%%%%%%%%%%%%%%%%%%%%%%%%%%%%%%%%%%%%%%%%%%%%%%%%%%

The plateau transitions of the quantum Hall effect (QHE) have retained interest over numerous decades \cite{vonKlitzing202040Effect,Girvin2019ModernPhysics, Son2021Three-dimensionalTransitions,Minkov2016HaldaneResonators,Ohgushi2000SpinFerromagnet,Haldane1988ModelAnomaly,Raghu2008AnalogsCrystals,Laughlin1981QuantizedDimensions,Tang2019Three-dimensionalZrTe5,Hannahs1989QuantumCrystal,Hohls2002DynamicalTransition,HASHIMOTO2008,DamHM11,HasCFS12}. At least four reasons come to mind that might explain this continuing attention: (i) the QHE offers a fairly accessible approach, both theoretically and experimentally, to studying the interplay of many-body interactions and disorder, due to its low-dimensionality, its ready realization in well-understood semiconductor platforms and its by now fairly accommodating magnetic-field requirements \cite{Parmentier2016QuantumMagnets,Novoselov2007Room-TemperatureGraphene,Cao2012QuantizedCarriers,Hill1998BulkEta-Mo4O11}. Next, (ii) the QHE exhibits the simplest of the topological phase transitions, serving both as a springboard into the field and a convenient benchmark case for the many advances in topological systems in the last decade \cite{Bernevig2006QuantumWells, Hasan2010TopologicalInsulators,Qi2008TopologicalInsulators, BernevigBA2013TopologicalSuperconductors,Rachel2018InteractingReview,Ozawa2019TopologicalPhotonics,Ding2022Non-HermitianGeometries}. Still, the QHE also retains some of its mysteries with (iii) ongoing interest in its microscopic mechanisms \cite{Ilani2004,Romer2021TheRegime,Weis2011MetrologyEffect} and the importance of interactions in both integer and fractional QHEs \cite{Oswald2023RevisionEffect,Chklovskii1992,Cooper1993a,OswaldPRB2017,OswR17,Halperin2020FractionalEffects} and (iv) the remaining discrepancies between experimental measurements and theoretical predictions \cite{Kramer2005RandomDimensions,Slevin2009CriticalTransition,Li2009ScalingModel}.

The precise value of the critical exponent governing the plateau-to-plateau transitions, even in the integer QHE, is a particularly intriguing such mystery \cite{Gruzberg2017GeometricallyTransitions}. While experimental results seem to have converged towards a value of $\nu = 2.38$ \cite{Li2009ScalingModel}, high-precision numerical studies have decisively shifted from earlier estimates, then in reasonable agreement with the experimental values, to a significantly higher value of $\nu= 2.59 \pm 0.01$ \cite{Slevin2009CriticalTransition,Puschmann2019IntegerLattice}.
This later increase comes from three independent improvements, namely (1) studies with increased system sizes, allowing the analysis to move ever closer to the transition point, (2) better theoretical modelling of the behaviour close to the transition with a more convincing treatment of irrelevant finite-size corrections and (3) an improvement in the statistics of the generated data. 
A similar such improvement of experimental data, which could be achieved by (1) lowering experimental temperature and (2) better control of experimental parameters, has not yet been undertaken, but might lead to a similarly increasing exponent.
Nevertheless, in the absence of improved experimental results one is drawn to evaluating other theoretical approaches. In a series of papers \cite{Klumper2019RandomGeometry,Conti2021GeometryTransitions}, following on from \cite{Gruzberg2017GeometricallyTransitions}, it was recently argued that a mix of classical and quantum networks can lead to a reduced estimate of $\nu$ to values again in agreement with experimental studies. Still, it remains unclear that such models can truly capture the QHE situation. 
Nevertheless, one should at least try and see if all the theoretical models formerly giving $\nu\sim 2.39$ can now be shown to consistently give $\sim 2.6$ when ($1$-$3$) are followed.

This program is what we present here for the case of the real-space renormalisation group (RSRG) to the Chalker-Coddington (CC) network model of the integer QHE, a method judiciously adapted from a similar RSRG for classical percolation \cite{Stauffer1991IntroductionTheory}. In the CC model, the method had previously been shown to give $\nu = 2.39\pm 0.01$ \cite{Cain2005}. This is better than what should have been expected since the RSRG in the classical percolation only produces the critical exponent of the percolation transition within $\sim 7\%$.
As we will show here, by increasing the number of samples $500$-fold, while also employing arbitrary-precision arithmetic \cite{WolframResearch2022Mathematica}, we now find a value of $\nu = 2.51 \pm 0.11$. 
In addition, we use the improved RSRG to also study the mixed problem of classical and quantum nodes proposed in \cite{Klumper2019RandomGeometry,Gruzberg2017GeometricallyTransitions} mentioned above. Sadly, we find that the problem is very sensitive to the geometry of the chosen renormalisation group (RG) unit. While fixed point distributions can be constructed and estimates of critical exponents can be found, these do not readily correspond to known universality classes.

\pagebreak
%%%%%%%%%%%%%%%%%%%%%%%%%%%%%%%%%%%%%%%%%%%%%%%%%%%%%%%%%%%%%%%%%%%
\section{Real-Space Renormalisation Group on the Chalker-Coddington Model}
%%%%%%%%%%%%%%%%%%%%%%%%%%%%%%%%%%%%%%%%%%%%%%%%%%%%%%%%%%%%%%%%%%%

%%%%%%%%%%%%%%%%%%%%%%%%%%%%%%%%%%%%%%%%%%%%%%%%%%%%%%%%%%%%%%%%%%%
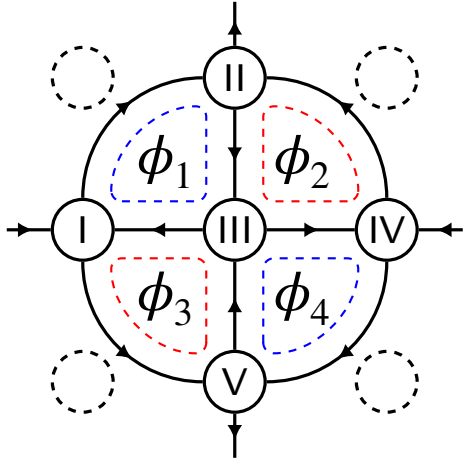
\begin{figure}[t]
\centering
\begin{tikzpicture}[scale=0.5,bend angle=45,decoration={
    markings,
    mark=at position 0.6 with {\arrow{Latex[length=2mm, width=2mm]}}}]
\tikzstyle{saddle}=[circle,very thick,draw=black,minimum size=8mm]   
    \node(three) at (0,0) [saddle] {};
    \node(one) at (-4,0) [saddle] {};
    \node(four) at (4,0) [saddle] {};
    \node(two) at (0,4) [saddle] {};
    \node(five) at (0,-4) [saddle] {};
    \node(nwempty) at (-4,4) [circle,very thick, dashed,draw=black,minimum size=8mm] {};
    \node(neempty) at (4,4) [circle,very thick, dashed,draw=black,minimum size=8mm] {};
    \node(seempty) at (4,-4) [circle,very thick, dashed,draw=black,minimum size=8mm] {};
    \node(swempty) at (-4,-4) [circle,very thick, dashed,draw=black,minimum size=8mm] {};
    \draw[postaction={decorate}, very thick] (one)  to[bend right] (five);
    \draw[postaction={decorate}, very thick] (one)  to[bend left] (two);
    \draw[postaction={decorate}, very thick] (four)  to[bend left] (five);
    \draw[postaction={decorate}, very thick] (four)  to[bend right] (two);
    \draw[postaction={decorate}, very thick] (three)  to (one);
    \draw[postaction={decorate}, very thick] (three)  to (four);
    \draw[postaction={decorate}, very thick] (five)  to (three);
    \draw[postaction={decorate}, very thick] (two)  to (three);
    \draw[postaction={decorate},very thick] (-6,0) to (one);
    \draw[postaction={decorate},very thick] (6,0) to (four);
    \draw[postaction={decorate},very thick] (two) to (0,6);
    \draw[postaction={decorate},very thick] (five) to (0,-6);
    \node[scale=1.6] at (one) {I};
    \node[scale=1.6] at (two) {II};
    \node[scale=1.6] at (three) {III};
    \node[scale=1.6] at (four) {IV};
    \node[scale=1.6] at (five) {V};
    \node[scale=2.4] at (1.8,1.8) {$\phi_2$};
    \node[scale=2.4] at (1.8,-1.8) {$\phi_4$};
    \node[scale=2.4] at (-1.8,1.8) {$\phi_1$};
    \node[scale=2.4] at (-1.8,-1.8) {$\phi_3$};
    \draw[rounded corners, thick, dashed, blue] (-3.25,0.75) -- (-0.75,0.75) -- (-0.75,3.25) to [bend right] cycle;
    \draw[rounded corners, thick, dashed, red] (3.25,0.75) -- (0.75,0.75) -- (0.75,3.25) to [bend left] cycle;
    \draw[rounded corners, thick, dashed, red] (-0.75,-3.25) -- (-0.75,-0.75) -- (-3.25,-0.75) to [bend right] cycle;
    \draw[rounded corners, thick, dashed, blue] (0.75,-3.25) -- (0.75,-0.75) -- (3.25,-0.75) to [bend left] cycle;
\end{tikzpicture}
\caption[Real-Space Renormalisation Group Unit Cell.]{The RSRG unit-cell of the CC model consisting of five saddle points. It is the smallest such construction which satisfies self-similarity with respect to edges. Nodes within the unit cell (circles) are labelled with roman numerals. Phases exist along each edge (arrowed lines) of the network, although in calculations these simplify to the four phases $\phi_1,\phi_2 ,\phi_3,\phi_4$, representing the phase accumulated upon traversing the edges associated with the dashed loop surrounding the label. The dashed circles represent ignored nodes.}
\label{fig:rgunitcell}
\end{figure}
%%%%%%%%%%%%%%%%%%%%%%%%%%%%%%%%%%%%%%%%%%%%%%%%%%%%%%%%%%%%%%%%%%%

%%%%%%%%%%%%%%%%%%%%%%%%%%%%%%%%%%%%%%%%%%%%%%%%%%%%%%%%%%%%%%%%%%%
\subsection{Scattering matrix approach to the QH transition}
\label{sec:CC-matrix}

The CC model describes the magnetic-field-induced chiral transport in the integer QHE via a 2D lattice populated with nodes representing saddle points in a continuous potential landscape. Electron wave packets can travel along directed equipotential paths -- clockwise or counterclockwise according to the direction of the magnetic field -- and scatter between such paths via tunelling across the saddlepoint nodes. Self-interference along the paths leads to localization while the tunnelling enhances electron transport. Taken together, the effects combine such that in summary the CC model exhibits a localisation-delocalisation-localisation transition at a single energy, hence modelling the plateau-to-plateau transitions of the QHE.
The CC model has been previously employed in many studies of integer QHE physics \cite{Chalker1988PercolationEffect,Kramer2005RandomDimensions,Zirnbauer2021MarginalTransition,Sedrakyan2003ActionEffect,Chalker1988ScalingEffect,Kramer2003LocalizationRegime,Evers1998SemiclassicalEffect}. In particular, when coupled with finite-size scaling, it can be used in transfer matrix studies to estimate the value of $\nu$ as discussed above \cite{Slevin2009CriticalTransition}.

% setting the scene, i.e. S matrix parametrizations
Mathematically, 
%, and edges connecting nodes representing equipotential paths providing electron transport between nodes. To analyse the scaling properties of the lattice, particularly the value of $\nu$, we utilise a real-space renormalisation group method. E
each saddle point is represented by a $2\times2$ matrix $\mathbf{S}$ connecting two incoming, $\vec{I}=(I, I')$, with two outgoing channels, $\vec{O}= (O, O')$, as $\vec{O}= \mathbf{S} \cdot \vec{I}$. Charge conservation is expressed via unitarity such that $|\vec{O}|^2= \vec{O}^\dagger \vec{O} = \vec{I}^\dagger \mathbf{S}^\dagger \mathbf{S} \vec{I} = | \vec{I} |^2$ with $\dagger$ representing a hermitian conjugate. The most general complex-valued  matrix $\mathbf{S}$ obeying these constraints is
\begin{equation}
\label{eq:smatrix-general}
\mathbf{S}
= \begin{bmatrix}
\alpha & \beta\\
-\beta^* \exp i \phi & \alpha^* \exp{i \phi}
\end{bmatrix}
\end{equation}
with $\alpha$, $\beta$ $\in \mathbb{C}$, $*$ the complex conjugate and $\phi$ a phase. 
Popular choices are
\begin{equation}
\label{eq:smatrix}
\mathbf{S}
= \begin{bmatrix}
t & r\\
r & -t
\end{bmatrix}
\hfil \text{and} \hfil
%    \label{eq:thetamatrix}
    \mathbf{S} = \begin{bmatrix}
        i\cos\theta & \sin\theta \\
        \sin\theta & i\cos\theta
    \end{bmatrix}.
\end{equation}
Here, 
$t$, $r$ $\in \mathbb{R}$ refer to \emph{transmission} and \emph{reflection amplitudes} in the first representation with $|t|^2+|r|^2 = 1$ \cite{Cain2001b} while the latter uses a single \emph{mixing angle}, $\theta$ \cite{Son2021Three-dimensionalTransitions}.

With reference to the potential landscape, an \emph{effective saddle point height}, $z$, can be defined by \cite{Galstyan1997LocalizationApproach}
\begin{equation}
    \cos^2\theta = t^2 = \frac{1}{1+e^z}, \quad 
    \sin^2\theta = r^2 = \frac{1}{1+e^{-z}} .
\end{equation}
In the following, we shall denote the distribution of these parameters over all saddle points as $P(t)$, $P(z)$ and $P(\theta)$. A further parameter, of perhaps more direct experimental relevance, is the dimensionless \emph{conductance} $G= t^2$ with $P(G)=P(t)/2t$. 
%Also, $P(z)=P(G) |dG/dz|=\frac{1}{4}\cosh^{-2}{(z/2)}P\left[ (e^{z} +1)^{-1} \right]$.
Furthermore, $P[t(z)]= P(z)|dt/dz|= e^{z} P[1/\sqrt{1+e^z}]/ 2 (1 + e^z)^{3/2} $ and $P[z(t)]= P(t) |dz/dt|= 2 P( \ln (-1+ 1/t^2) )/ (t - t^3)$.

%%%%%%%%%%%%%%%%%%%%%%%%%%%%%%%%%%%%%%%%%%%%%%%%%%%%%%%%%%%%%%%%%%%
\subsection{RG determination of the fixed point distributions at the QH transition}
\label{sec:CC-FP}

% defining the RG
Transfer matrix and diagonalization studies of the CC model usually build up quasi-1D strips or 2D square lattices of saddle points and then proceed to compute localization lengths, participation numbers, etc. \cite{Lee1993QuantumEffect,LSchweitzer1984MagneticSystems}. The RSRG approach, on the other hand, works by considering a small subset of such a lattice structure by constructing an RG unit from several neighbouring nodes \cite{Galstyan1997LocalizationApproach}. For the RG unit, we assemble the $\mathbf{S}$ matrices of each participating node into a \emph{combined} $\mathbf{S}$ matrix and remove all unwanted connections to superfluous in- and outgoing channels such that we are only left with again two incoming and two outgoing channels. In figure \ref{fig:rgunitcell}, we show this RG unit graphically. We note that other RG units are possible, but none have thus far been shown to yield better results \cite{Assi2019ADisorder,Cain2005} than the $5$-node RG unit depicted in figure \ref{fig:rgunitcell}. 
We emphasize that in drawing the figure, we have 
used the fact that the phases $\phi_j$ along the four closed loops $j$ add.
The five $2 \times 2$ $\mathbf{S}$ matrices combine into a single $10 \times 10$ matrix equation, previously shown in Refs.\ \cite{Cain2005,Assi2019ADisorder} in the $t$, $r$ representation. Using $\theta$, we find equation \ref{eq:thetatransform}.
%\pagebreak
%
In principle, starting with the five saddle points described by mixing angles $\theta_i$, $i=1, \ldots, 5$ and the phases of the four closed loops $\phi_1, \phi_2, \phi_3, \phi_4$, we can compute the effective mixing angle $\theta'$ of the super-saddle point.

% determining the fix point distribution
 
We are now ready to proceed with the RG process itself. We first need to construct a starting distribution $P_1$ of parameters. Then application of the RG for $k$ RG generations will allow us to find the distribution of super-parameters $P_k$ \cite{Galstyan1997LocalizationApproach}. At criticality, we expect that $P_k \rightarrow P_\text{FP}$ for $k \rightarrow \infty$ with $P_\text{FP}$ denoting the unstable fixed point (FP) distribution \cite{Cain2005}.\footnote{Starting from a distribution too asymmetric or to far away from the $P_\text{FP}$, the RG flow will tend towards the stable classical FPs, e.g.\ $P(G)=\delta(G)$ or $P(G)=\delta(G-1)$.} 
%\pagebreak

%
While this procedure can in principle be done using any of the $t$, $z$, or $\theta$ representations, here we choose a combination of $t$ and $z$ representations for convenience \cite{Cain2001b}. 
We start with $P_1(t)=2t$, equivalent to $P_1(z)= \frac{e^z}{(e^z+1)^2}$ (and $P_1(G)=1$), and generate $P_2(t)$. The phases $\phi_i$ are chosen randomly in $[0, 2\pi]$ to model spatial variation among equipotentials.
\begin{strip}
\begin{equation}\label{eq:thetatransform}
\cos \theta^{\prime}=
\stretchleftright{|}
{
\frac{\splitfrac{-\cos \theta_1\left[ e^{i \phi_4} \cos \theta_3 \cos \theta_4+i \cos \theta_5\left(e^{i \phi_3} \sin \theta_2 \sin \theta_3 \sin \theta_4-1\right)\right]-e^{i \phi_1} \cos \theta_2 \cos \theta_3 \cos \theta_5 }{ -i e^{i\left(\phi_1+\phi_4\right)} \cos \theta_4 \cos \theta_2+i e^{i\left(\phi_1+\phi_4-\phi_2\right)} \cos \theta_2 \cos \theta_4 \sin \theta_1 \sin \theta_3 \sin \theta_5}}{\splitfrac{e^{i \phi_4} \cos \theta_3 \cos \theta_4 \cos \theta_5+e^{i \phi_1} \cos \theta_1 \cos \theta_2
\left\{ \cos \theta_3+i e^{i \phi_4} \cos \theta_4 \cos \theta_5 \right.}{\left. +i\left[ e^{i \phi_3} \sin \theta_2 \sin \theta_3 \sin \theta_4+e^{i \phi_2} \sin \theta_1\left( \sin \theta_3-e^{i \phi_3} \sin \theta_2 \sin \theta_4 \right) \sin \theta_5 - 1\right] \right\}}}
}{|}.
\end{equation}
\end{strip}
We then transform each $t$ value to $z$ and find $P_2(z)$. \citet{Cain2001b} have shown that the approach to the fixed point results in a sequence of symmetric distributions such that $P_k(z)=P_k(-z)$ with $\langle P_k(z) \rangle \approx 0$. Conversely, the symmetry can be enforced for each $P_k(z)$ and a symmetrized $P_k(z)$ with $\langle P_k(z) \rangle = 0$ constructed following the transformation relations given at the end of section \ref{sec:CC-matrix}. This stabilizes the approach to the fixed point distributions $P_\text{FP}(t)$, $P_\text{FP}(z)$, etc.
%
%We can select as $P_0(z)$ a symmetric function $P_0(z)=P_0(-z)$ with $\langle P_0(z) \rangle=0$ and find $P_k(z) \rightarrow P_\text{FP}(z)$ with $P_\text{FP}(z)$ shaped similar to a Gaussian. 
%%%%%%%%%%%%%%%%%%%%%%%%%%%%%%%%%%%%%%%%%%%%%%%%%%%%%%%%%%%%%%%%%%%
\begin{figure*}[t]
    \centering
    \begin{subfigure}{0.48\textwidth}
        \includegraphics[width=0.99\columnwidth]{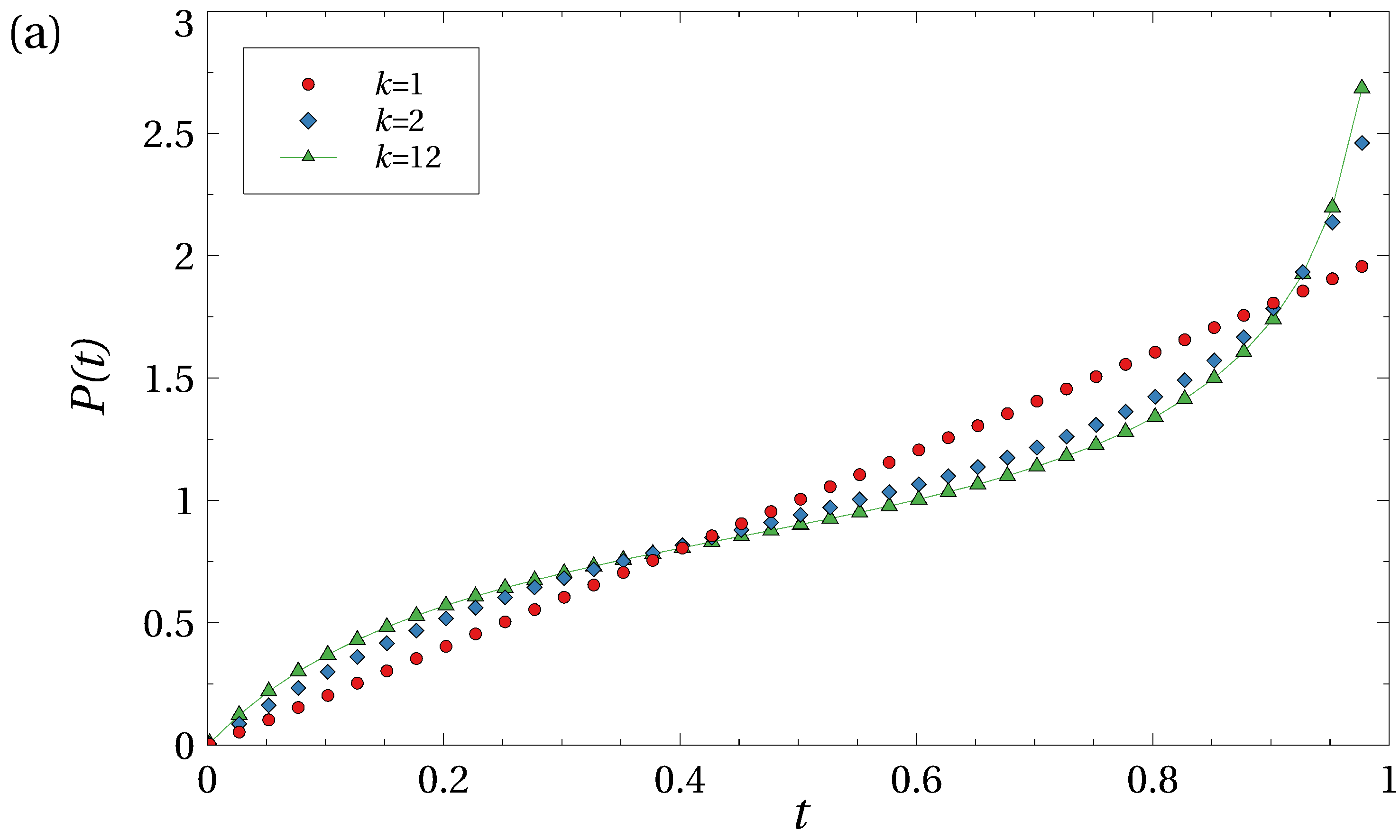}
    \end{subfigure}
    \begin{subfigure}{0.48\textwidth}
        \includegraphics[width=0.99\columnwidth]{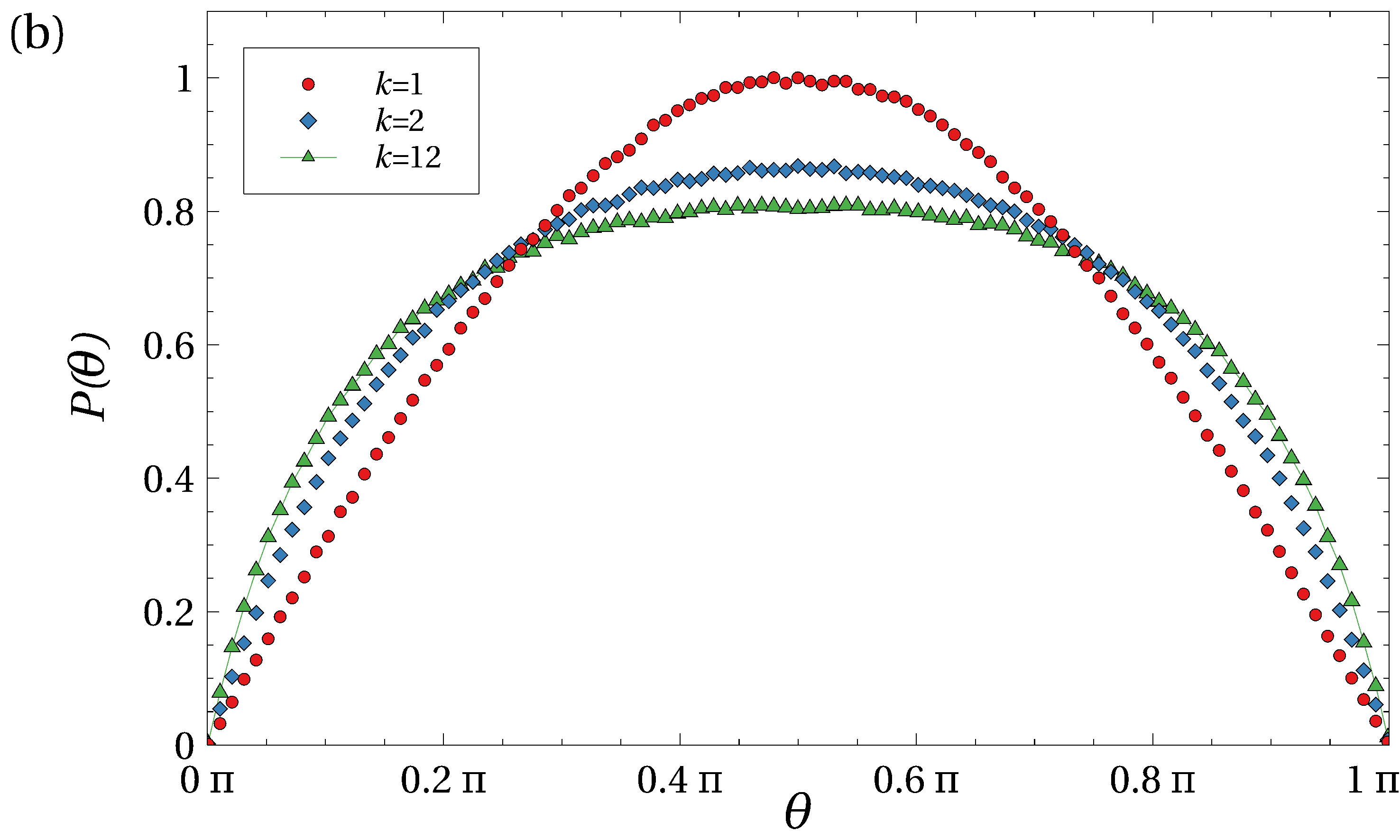}
    \end{subfigure}
    \begin{subfigure}{0.48\textwidth}
        \includegraphics[width=0.99\columnwidth]{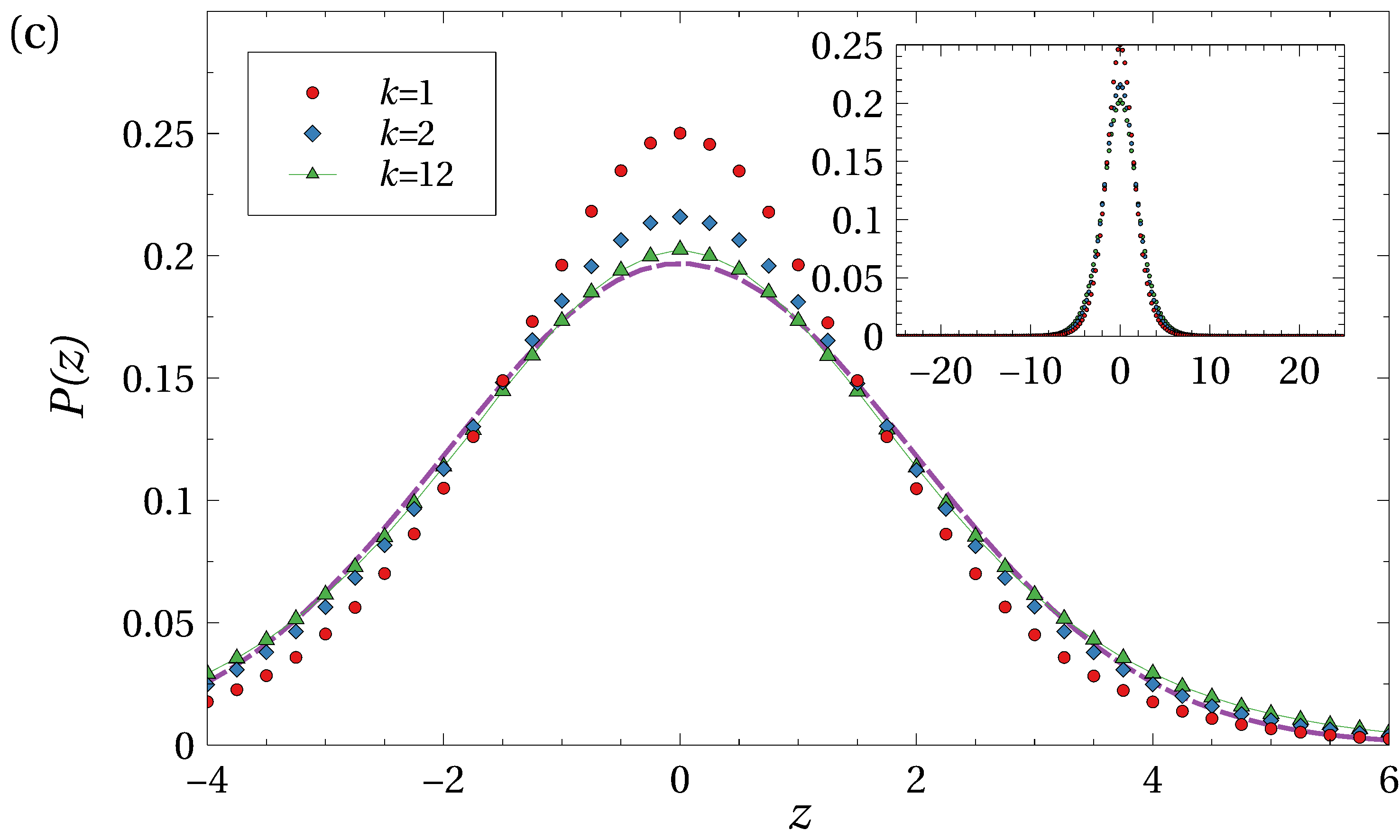}
    \end{subfigure}
    \begin{subfigure}{0.48\textwidth}
        \includegraphics[width=0.99\columnwidth]{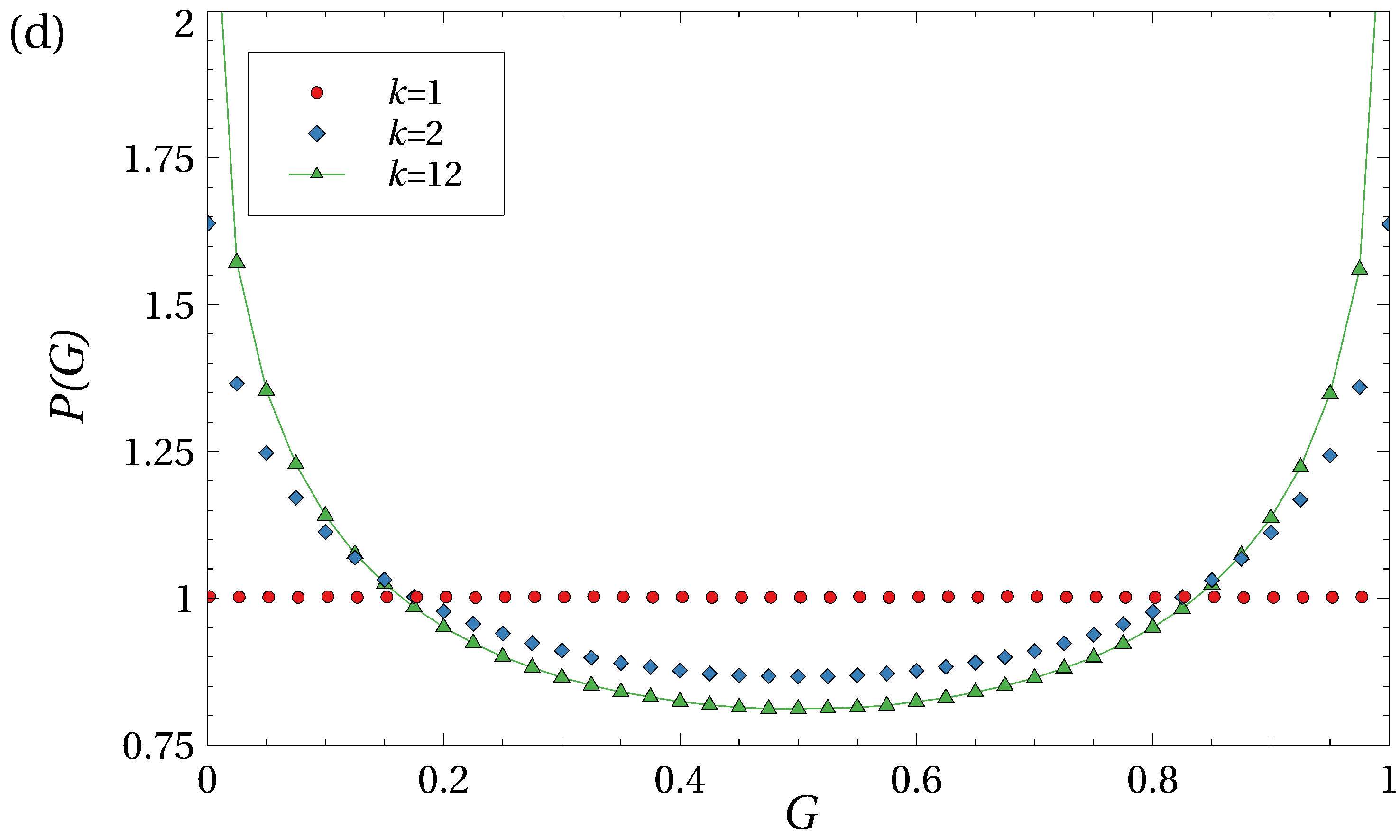}
    \end{subfigure}
    \caption{Evolution of the distribution $P$ of elements of the scattering matrix for RSRG steps $k=1, 2$ and  reaching close to convergence at $k=12$ in the (a) $P(t)$, (b) $P(\theta)$, (c) $P(z)$ and (d) $P(G)$ representations. While in (a+b+d) the full range of $t$, $\theta$ and $G$ is displayed, panel (c) shows a smaller $z$ range for clarity while the full range $z\in [-25,25]$ is given in the inset. The thin solid line in all panels is a guide to the eye for $P_{12}(z)$. The dashed line in panel (c) shows a Gaussian fit to $P_{12}(z)$ for all $z\in [-25,25]$. } 
    %converged $P_\text{FP}(z)$ at $k=???$.}
    \label{fig:fixpointdistributions}
\end{figure*}
%%%%%%%%%%%%%%%%%%%%%%%%%%%%%%%%%%%%%%%%%%%%%%%%%%%%%%%%%%%%%%%%%%%
%We can select as $P_0(z)$ a symmetric function $P_0(z)=P_0(-z)$ with $\langle P_0(z) \rangle=0$ and find $P_k(z) \rightarrow P_\text{FP}(z)$ with $P_\text{FP}(z)$ shaped similar to a Gaussian.

We implement this procedure and generate at least $5\times 10^9$ $z$ values for each RG generation $k$. This is about $500$ times more than in previous such RG studies \cite{Cain2003c}. Since the generated super-distributions will be no longer comprised of simple functions, we employ the rejection method \cite{Press2007} to generate appropriate random numbers. Also, in the version implemented for this work, we use \textsc{Mathematica}'s arbitrary-precision arithmetic when evaluating Eq.\ \eqref{eq:thetatransform} (in the $t$ representation \cite{Cain2001b}). 
This reduces inevitable rounding errors when computing $t$ and $z$ values with usual compiler-based double precision arithmetic, allowing us to keep accurate track of $z$ values ranging from $z_{<}=-25$ ($1 - t_{>}\approx 10^{-11})$ to $z_{>}=25$ ($t_{<}\approx 10^{-11})$.

In figure \ref{fig:fixpointdistributions}, we show the resulting FP distributions for (a) $P(t)$, (b) $P(\theta)$, (c) $P(z)$ and (d) $P(G)$. We note that the spread of $P_\text{FP}(z)$ around $z=0$ reflects the distribution of the disorder landscape. 
%
% We note that the symmetry $P_\text{FP}(z)=P_\text{FP}(-z)$ is retained at the FP. Hence this property can be imposed as a stabilizing condition during the RG procedure, preventing the build-up of corrections towards the stable, classical FPs. The resulting symmetry-stabilized FP distributions are also indicated in figure \ref{fig:fixpointdistributions}.
%
A fit to a Gaussian for $z\in [-25,25]$ can be obtained via a usual $\chi^2$ minimization taking into account the uncertainties of the $P_\text{FP}(z)$ values. However, a perfect fit is only achieved when we increase the uncertainties $8$-fold. Nevertheless, the coefficient of determination is $R^2 > 0.99$ with $\langle P(z) \rangle= (0\pm 2)\times 10^{-4}$ and standard deviation $\sigma= 2.17$. 
The fitted Gaussian curve is shown alongside the data in Figure \ref{fig:fixpointdistributions}(c). We conclude that a Gaussian approximation is suitable, in particular for determining the maximum value of successive distributions with a restricted $z$ range, and we use it below in determining the mean shift in saddle point heights.

%%%%%%%%%%%%%%%%%%%%%%%%%%%%%%%%%%%%%%%%%%%%%%%%%%%%%%%%%%%%%%%%%%%
\subsection{Determination of an improved critical exponent}
\label{sec:CC-critexp}

% computing the critical exponent

\begin{figure}[tb]
    \centering
    \includegraphics[width=0.99\columnwidth]{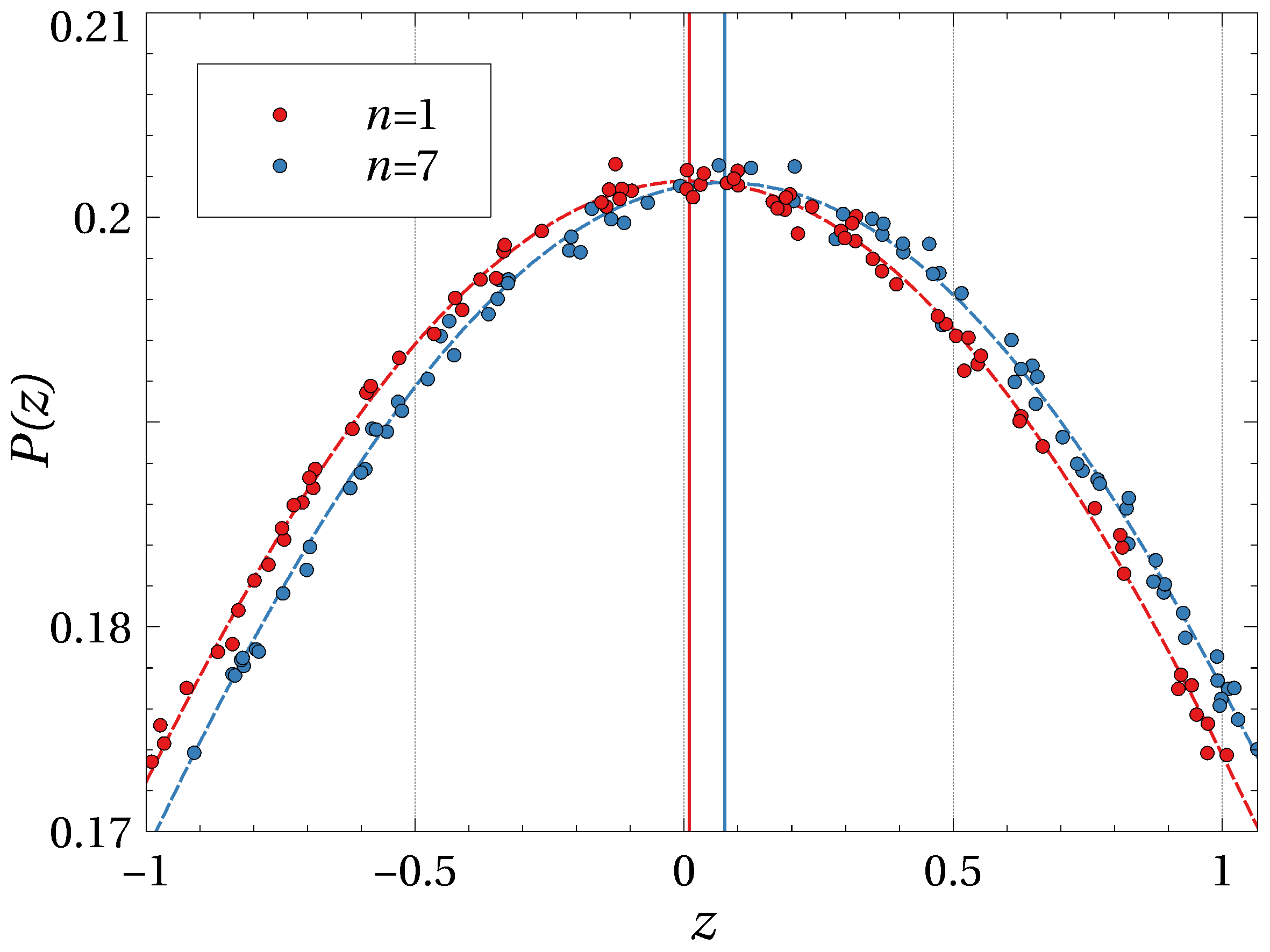}
    \caption{Example of a shift of $P_{\text{FP}}(z)$ when perturbing with $z_0=0.1$. The data shown corresponds to $n=1$ and $n=7$, the lines indicate Gaussian fits and the two vertical lines show the obtained estimates for $z_\text{max}$. These two data values have been used when plotting Fig.\ \ref{fig:z0-rayplot}. Vertical dashed grid lines mark $z=\mp 0.5$ and $1$.}
    \label{fig:z0-driftplot}
\end{figure}

\begin{figure}[bt]
    \centering
    \includegraphics[width=0.99\columnwidth]{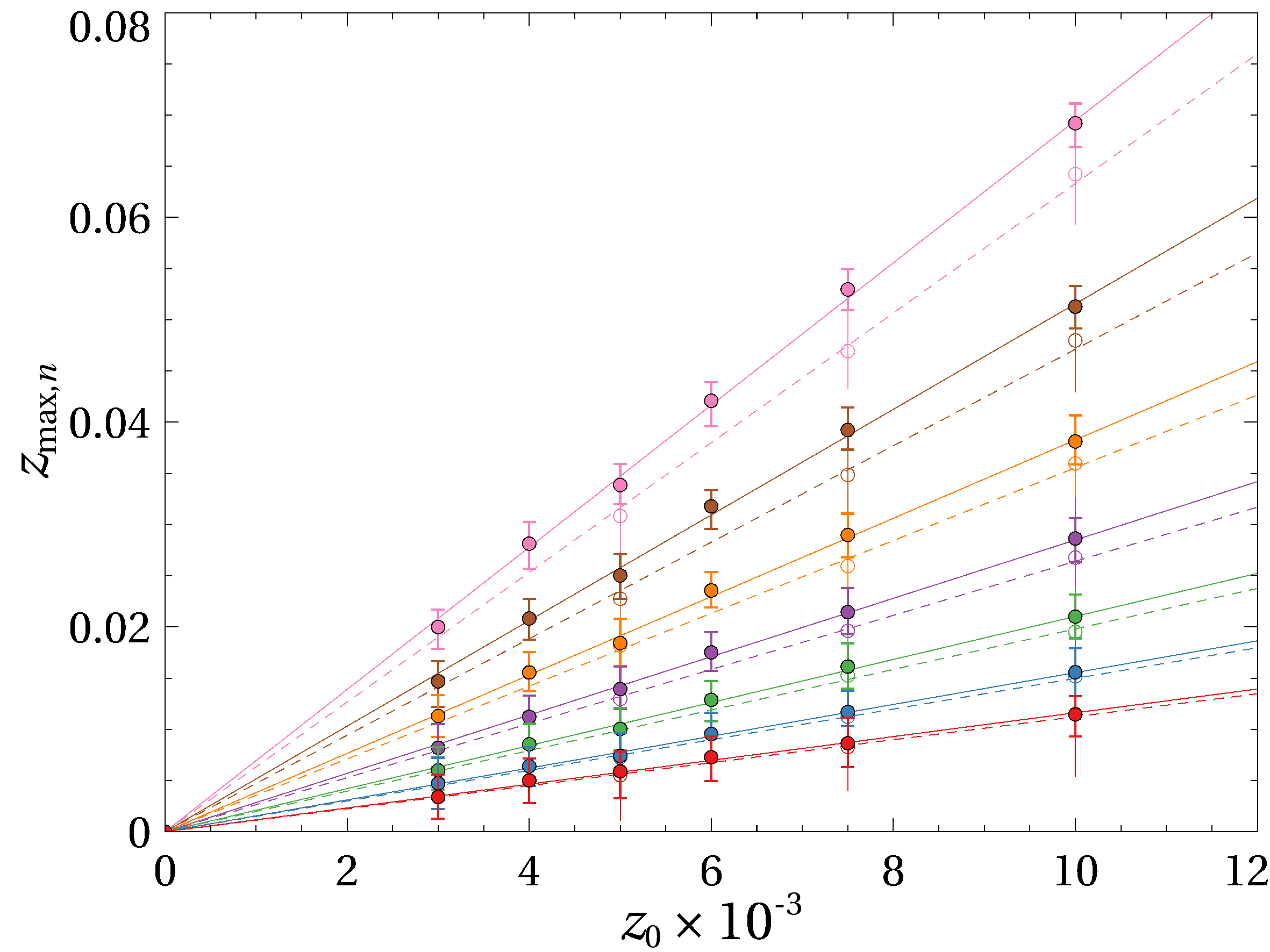}
    \caption{Linear dependence of $z_\text{max}$ on $z_0$ for six different starting values of $z_0$s. Solid lines and markers indicate data without inclusion of the mixed model, whereas dashed lines with unfilled markers indicate data gathered from the mixed model with $p=0.01$. Each set of lines corresponds to RG levels $n= 1, \ldots, 7$ from bottom to top. 
    The error bars for the CC model represent the highest and lowest values attained for five independent RSRG calculations with $10^9$ $z$ values each. The average value used henceforth is the mean of all $5$ $z_\text{max}$ values. 
    For the mixed disorder model, the error is calculated from ten independent RSRG calculations for $10^8$ $z$ values.
    Each $z_\text{max}$ datum is obtained via Gaussian fits as shown in Fig.\ \ref{fig:z0-driftplot}. Lines represent the fits for $z_n/z_0$ according to Eq.\ \eqref{eq:nuequation}.}
    \label{fig:z0-rayplot}
\end{figure}

\begin{figure}[t]
\centering
\includegraphics[width=0.99\columnwidth]{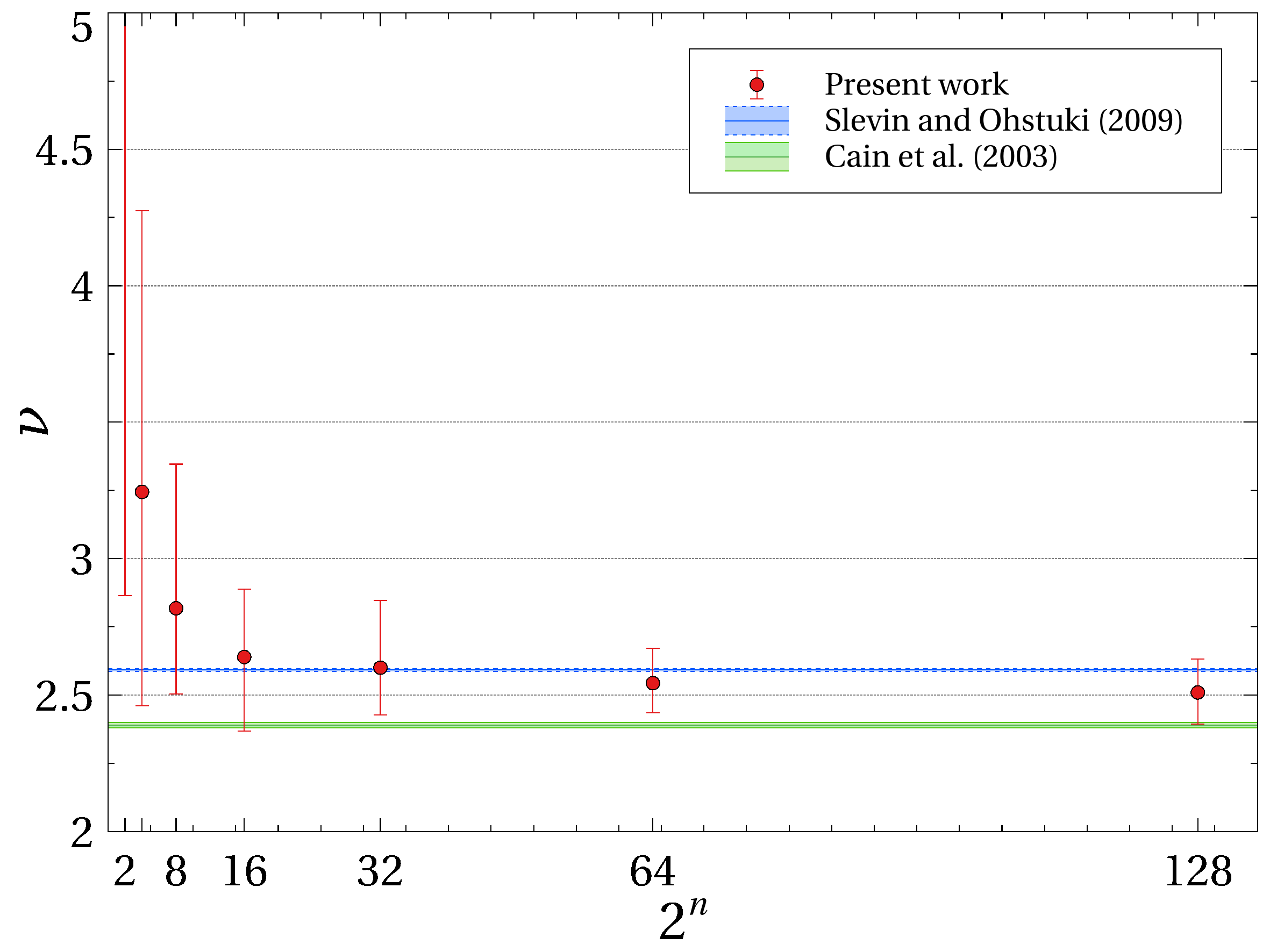}
\caption[The Critical Exponent of the Correlation Length.]{A plot of $\nu$ as a function of system size $2^n$, where $n$ is the renormalisation step number. Seven RG steps are plotted. Each data point represents three values of $z_0$ averaged over 10 instances of our method, each instance consisting of $5\times10^9$ samples. We find the final value of $\nu$ to be $2.51\substack{+0.11\\-0.11}$. The blue and green regions indicate reference values and associated errors found for $\nu$ previously \cite{Slevin2009CriticalTransition,Cain2003a}, as indicated in the legend.}
\label{fig:nuplot}
\end{figure}

To determine the critical exponent $\nu$, we next perturb the FP distribution by a constant shift $z_0$, i.e.\ $P_\text{FP}(z) \rightarrow P_\text{FP}(z-z_0)$. We then observe how the perturbed distribution drifts away from criticality by monitoring the deviation of its maximum at $z_\text{max}$ from $0$ under repeated (unsymmetrized) RG steps \cite{Cain2001b}. Only the largest $4\%$ of $z$ values, corresponding approximately to all $z\in [z_0-1, z_0+1]$, are used to determine $z_\text{max}$ by fitting to a Gaussian as shown in figure \ref{fig:z0-driftplot}. 
%%%%%%%%%%%%%%%%%%%%%%%%%%%%%%%%%%%%%%%%%%%%%%%%%%%%%%%%%%%%%%%%%%%
%%%%%%%%%%%%%%%%%%%%%%%%%%%%%%%%%%%%%%%%%%%%%%%%%%%%%%%%%%%%%%%%%%%
After $n$ such RG steps, we find that the position of the maximum follows $z_\text{max}(n) \propto z_0$ with good accuracy as detailed in figure \ref{fig:z0-rayplot}. 
%%%%%%%%%%%%%%%%%%%%%%%%%%%%%%%%%%%%%%%%%%%%%%%%%%%%%%%%%%%%%%%%%%%
%%%%%%%%%%%%%%%%%%%%%%%%%%%%%%%%%%%%%%%%%%%%%%%%%%%%%%%%%%%%%%%%%%%
The critical exponent can then be computed via \cite{Galstyan1997LocalizationApproach}
\begin{equation}
\label{eq:nuequation}
    \nu = \frac{\ln 2^n}{\ln\left(\frac{z_\text{max}(n)}{z_0}\right)}
\end{equation}
and should converge for $n \rightarrow \infty$.
As presented in figure \ref{fig:nuplot}, we find $\nu \approx 2.51\substack{+0.11\\-0.11}$ for $n=7$.
%%%%%%%%%%%%%%%%%%%%%%%%%%%%%%%%%%%%%%%%%%%%%%%%%%%%%%%%%%%%%%%%%%%
%%%%%%%%%%%%%%%%%%%%%%%%%%%%%%%%%%%%%%%%%%%%%%%%%%%%%%%%%%%%%%%%%%%
In obtaining this increased $\nu$ value, the enhanced statistics of $P(z)$ allows us to study initial perturbations $z_0= 0.003, 0.004, 0.005, 0.006, 0.0075$, and $0.01$. These values are ten times smaller than used previously, hence increasing the accuracy of the determination for $\nu$. With $\nu= 2.593 \substack{+0.005\\-0.006}$ the accepted value from transfer matrix calculations \cite{Slevin2009CriticalTransition}, we find that our new result only deviates by $3.2\%$. This is of course very good when compared to the $7\%$ accuracy in classical percolation \cite{Stauffer1991IntroductionTheory}.

%%%%%%%%%%%%%%%%%%%%%%%%%%%%%%%%%%%%%%%%%%%%%%%%%%%%%%%%%%%%%%%%%%%
\section{Mixing classical and quantum percolation}
%%%%%%%%%%%%%%%%%%%%%%%%%%%%%%%%%%%%%%%%%%%%%%%%%%%%%%%%%%%%%%%%%%%

While our new $\nu$ is now in better agreement with the high-precision estimates of \citet{Slevin2009CriticalTransition} and \citet{Puschmann2019IntegerLattice}, we have similarly increased the distance to the lower experimental estimates \cite{Li2009ScalingModel,Hohls2002DynamicalTransition}. 
Work by \citet{Nuding2015} and others \cite{Gruzberg2017GeometricallyTransitions, Klumper2019RandomGeometry} argues that this discrepancy could lie in the highly irregular arrangement of experimental scattering nodes which is nevertheless modelled on a square lattice in the CC model. They suggest that a mixed model, in which some nodes are left in the CC configuration, with $t$ and $r$ values, while other nodes are chosen fully open ($t=1$, $r=0$) or fully closed ($t=0$, $r=1$) with probability $p$, could lower the value of $\nu$ to be in better agreement with the experiments.

%%%%%%%%%%%%%%%%%%%%%%%%%%%%%%%%%%%%%%%%%%%%%%%%%%%%%%%%%%%%%%%%%%%
\subsection{Fixed point distributions for the mixed model}
\label{sec:mixed-FP}

This type of mixed disorder, termed "geometric disorder" by \citet{Gruzberg2017GeometricallyTransitions}, can be implemented into the RSRG method by intentionally modifying $t$ values of the saddle points within the distributions. For distinction from the previous situation, we shall denote the resulting distributions as $Q_p(t)$, $Q_p(z)$, etc. With this notation, $P(z)=Q_0(z)$. 
Upon every $k$th RG transformation each saddle point in the RG unit cell becomes, with probability $p$, either entirely transmitting ($t=1$) or entirely reflecting ($t=0$). Or, with probability $1-2p$, $p\leq 1/2$, the saddle point transmission amplitude remains a random number $\in [0,1]$ selected from the previous distribution $Q_{p,k-1}(t)$, as before.
But simply setting $t$ values to $0$ (i.e.\ $\theta=\pi$) in the analytic equation determining the renormalised $t'$ (i.e.\ Eq.\ \eqref{eq:thetatransform}) can result in a vanishing denominator. 
In terms of $z$, the $t=0$ and $1$ values correspond to $z=\pm \infty$, respectively. Hence we see a build-up of large $\pm z \approx 34$ values in $Q_{p,k+1}(z)$, which is increasing as $k$ increases.
Clearly, if we were to use such distributions in our RG procedure, we would effectively double-count the influx of the mixed disorder \cite{Assi2019ADisorder}. We avoid this by neglecting all accumulated $z$ values beyond $z_\text{+,--}=\pm 25$ when computing $Q_{p,k+1}(z)$ in each RG step. 
%In terms of $t$, these bounds equate to $t_\text{--}=1.39 \times 10^{-11}$ and $t_\text{+}=1-1.39 \times 10^{-11}= 0.999999999986\ldots$ . 
Overall, the proportion of such events, $2\int_{-\infty}^{z_-} Q_{p,k}(z) dz$, remains less than $10^{-5}$ as we try to approach a new $Q_{p,\text{FP}}(z)$.
%%%%%%%%%%%%%%%%%%%%%%%%%%%%%%%%%%%%%%%%%%%%%%%%%%%%%%%%%%%%%%%%%%%
\begin{figure}
    \centering
    \includegraphics[width=0.99\columnwidth]{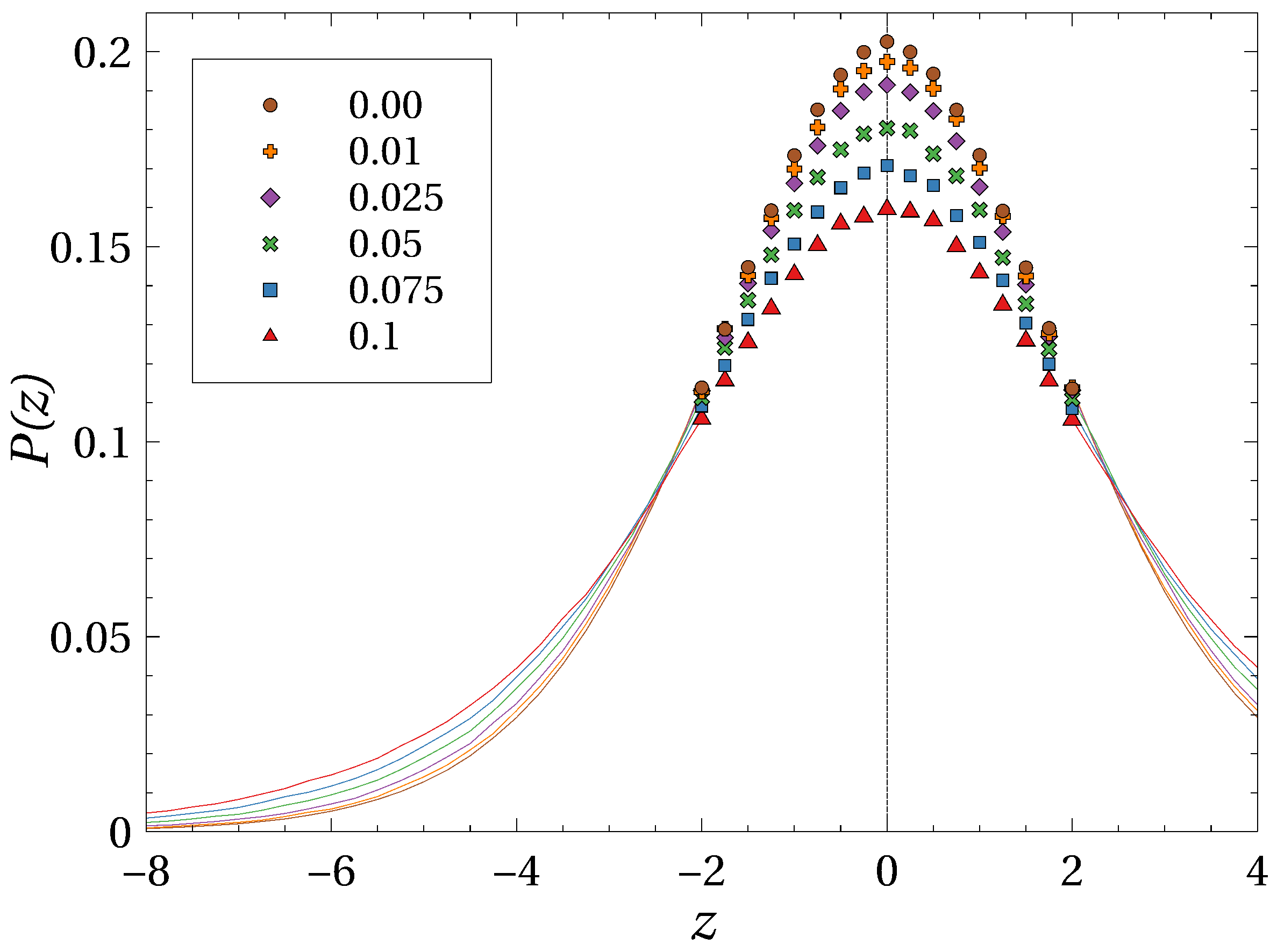}
    \caption{Fixed point distributions $Q_0(z)$ ($=P(z)$), $Q_{0.01}(z)$, $\ldots$, $Q_{0.1}(z)$ obtained with $1\times 10^9$ RSRG evaluations. The values around $z=0$ are indicated by symbols while we use lines otherwise. We also only show a small $z$-region $\in [-25,25]$ for clarity. In calculations, we use the full $z$-range as displayed in the inset of Fig.\ \ref{fig:fixpointdistributions}(c).}
    \label{fig:gdzdists}
\end{figure}
%%%%%%%%%%%%%%%%%%%%%%%%%%%%%%%%%%%%%%%%%%%%%%%%%%%%%%%%%%%%%%%%%%%

%%%%%%%%%%%%%%%%%%%%%%%%%%%%%%%%%%%%%%%%%%%%%%%%%%%%%%%%%%%%%%%%%%%
%p-dependence
In figure \ref{fig:gdzdists} we show that we can indeed find this new FP distribution for the case of mixed disorder for various $p$ values. The fixed point distributions are again of roughly Gaussian shape with (e.g.) $\langle Q_{0.1}(z)\rangle = (0\pm 2)\times 10^{-4}$ and $\Delta_p = | Q_{p,\text{FP}}(z) - Q_\text{Gau\ss}(z) | /N = 1,1,2,2,3$ ($\times 10^{-3}$) for $p=0.01$, $0.025$, $0.05$, $0.075$, $0.1$, respectively, where $N$ denotes the number of bins used in constructing the $Q(z)$ distributions. In relation to $P_\text{FP}$, we note that the height-to-width ratio of $Q_{p=0.1, \text{FP}}(z)$ is about $0.62$ times that of $Q_{p=0.0, \text{FP}}(z)$. The fixed point distributions with $p>0$ have significantly longer $z$-tails, corresponding to a more rugged saddle point height landscape.

\begin{figure}
    \centering
    \begin{subfigure}{0.99\columnwidth}
        \includegraphics[width=0.99\columnwidth]{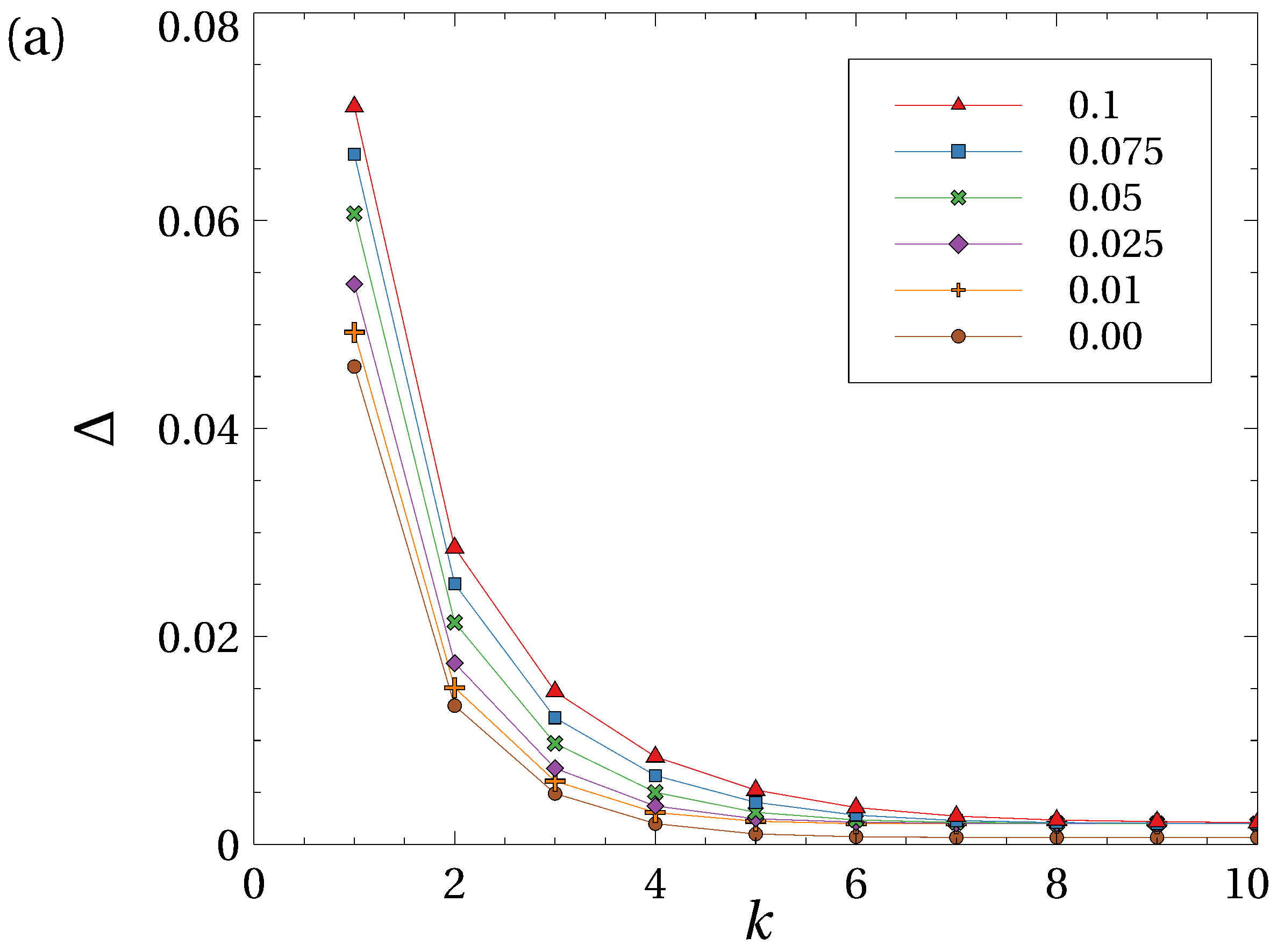}
    \end{subfigure}
    \begin{subfigure}{0.99\columnwidth}
        \includegraphics[width=0.99\columnwidth]{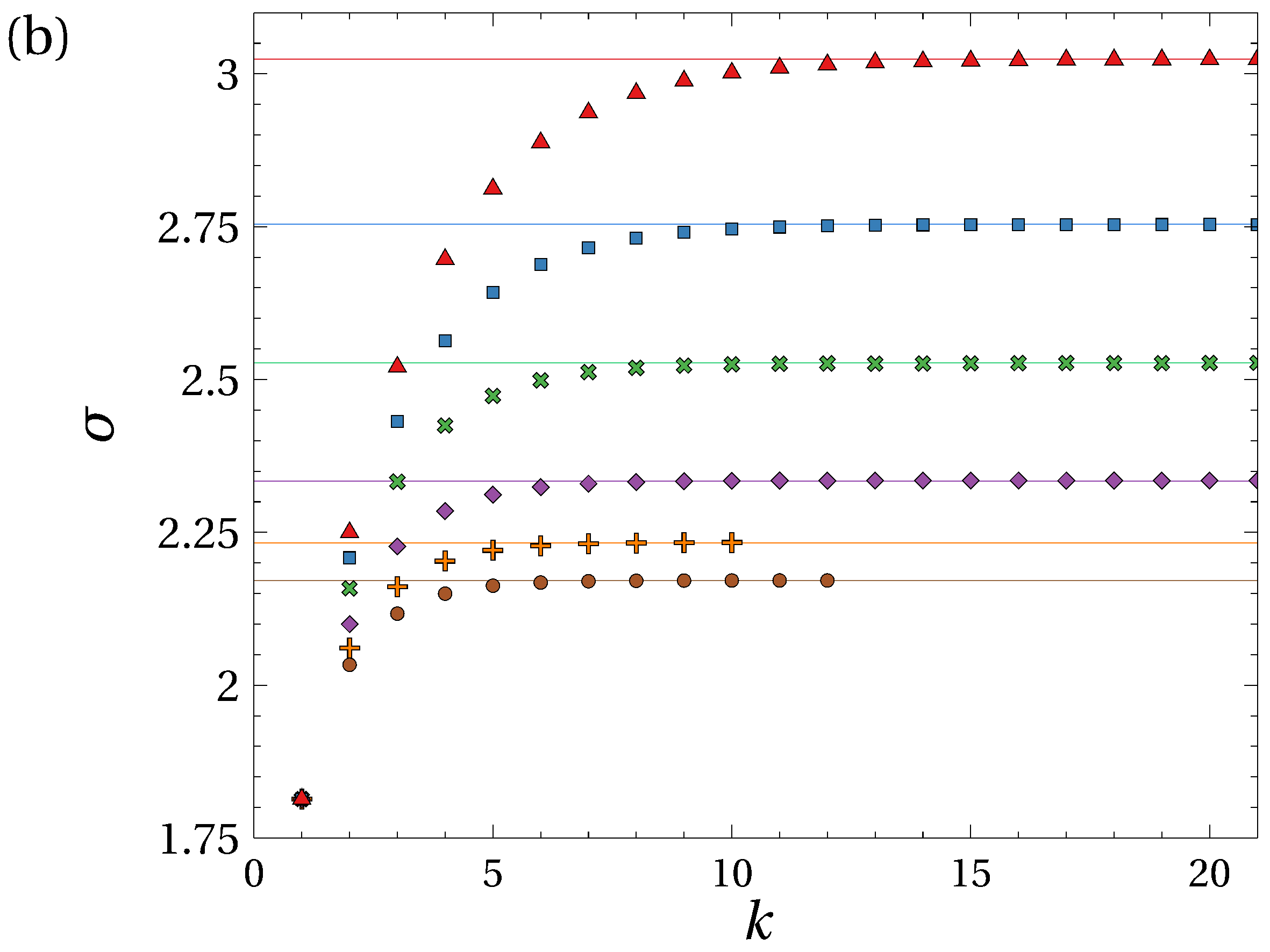}
    \end{subfigure}
    \caption{A comparison of key parameters indicating convergence towards a fixed point distribution $Q_{p,\text{FP}}$ for multiple values of $p$. 
    (a) shows the squared error between successive distributions $\Delta_p(k) = \int|Q_{k+1}(z)-Q_k(z)|^2 \mathrm{d}z$ across $20$ RG step iterations. 
    (b) shows $\sigma_p(k)$ values as the number of RG steps increases. 
    Plotting in (b) is consistent with the key in (a), but reversed to the key in figure \ref{fig:gdzdists} to follow the top-to-bottom order of the data. 
    For all values of mixed disorder proportion $p$, the $Q_p(z)$ distribution converges. As $p$ increases, more renormalisation steps are required for the distributions to converge. Lines in (a) are guides to the eye only while in (b) they indicate the converged values.}
    \label{fig:convergence}
\end{figure}

%The consecutive application of the renormalisation transformation and a symmetrisation procedure between steps successfully results in an FP distribution when varying the mixed disorder proportion $p$. 
The introduction of mixed disorder has no effect on the mean value of the $Q_{p, \text{FP}}(z)$ distribution, as random application of mixed disorder doesn't bias the effective saddle point heights in either direction. However, the introduction of $p$ notably increases the amount of steps required to converge towards an FP distribution. This is shown in figure \ref{fig:convergence} with $\Delta_p$ plotted against RG iteration number $k$ for varying values of $p$. Additionally, the shape of the FP distribution changes depending on $p$, as shown by the change in standard deviation $\sigma_p$ converged upon for different $p$ values also in figure \ref{fig:convergence}. For larger $p$, we note a consistent increase in $\sigma_p$ for each $Q_{p,\text{FP}}$.
%%%%%%%%%%%%%%%%%%%%%%%%%%%%%%%%%%%%%%%%%%%%%%%%%%%%%%%%%%%%%%%%%%%
%%%%%%%%%%%%%%%%%%%%%%%%%%%%%%%%%%%%%%%%%%%%%%%%%%%%%%%%%%%%%%%%%%%

%%%%%%%%%%%%%%%%%%%%%%%%%%%%%%%%%%%%%%%%%%%%%%%%%%%%%%%%%%%%%%%%%%%
\subsection{Critical exponent for the mixed model}
\label{sec:latentGD}
\label{sec:midex-critexp}

\begin{figure}[t]
\centering
    \includegraphics[width=0.99\columnwidth]{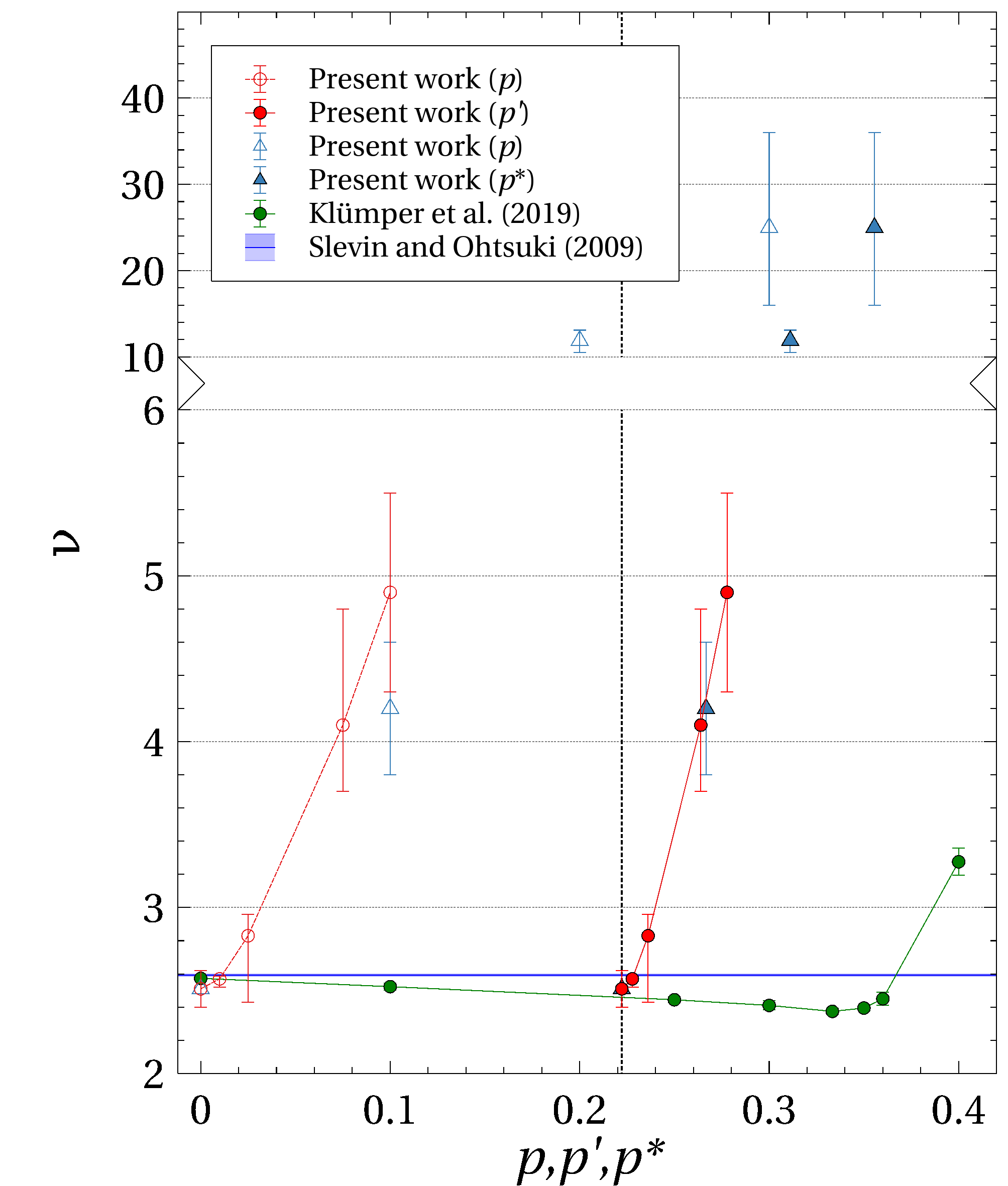}
\caption{Obtained estimates for $\nu$ compared with computed $\nu$ values from Ref.\ \cite{Klumper2019RandomGeometry} for varying proportions of mixed disorder. Circles distinguish our results with all nodes affected by mixing, from triangles, which represent the data from the unit cell with the central node unaffected by the mixed disorder. Data transformed to $p'$ and $p^*$ are shown with filled-in markers ($\circbox1{black}$, $\trianbox1{black}$), whereas original $p$ value markers have no fill ($\circbox0{black}$, $\trianbox0{black}$). Lines connecting $\nu$ estimates are guides to the eye only. The vertical dashed line indicates the horizontal probability value $2/9$.} 
%for $\nu$ using the RSRG unit cell with all nodes affected by mixing. (b) shows our results for $\nu$ when the central node in the RSRG unit cell is unaffected by the mixed disorder. Data points are connected by lines as a visual guide only.}
\label{fig:gdcomparison}
\end{figure}

% determination of critical exponents for various p values
After having determined the FP distributions $Q_{p,\text{FP}}(z)$, we shift these as before by a small $z_0$ to determine the RG flow of the position on the maxima $z_{p,\text{max}}$. %as shown in figure \ref{fig:z0-driftplot}. 
As indicated in figure \ref{fig:z0-rayplot}, we find good linearity and can hence proceed to determine, for each $p$, a corresponding critical exponent of the mixed model, $\nu_p$ via equation \eqref{eq:nuequation}.
The results are astounding: the values of $\nu_p$ increase monotonically from $\nu_{p=0}=2.51$ to $\nu_{p=1/3}=4.9$ as shown in Table \ref{tab:gdresults}. Particularly the last number seems to indicate a behaviour at the QHE plateau-plateau transitions not found experimentally \cite{Koch1991ExperimentsConditions}.
% larger values are in agreement with Sedrakyan
An increased value of $\nu$ up to $\nu_{p=0.4}=3.276$ was indeed observed in \citet{Klumper2019RandomGeometry}. But the overall quantitative agreement of our results with these previous results across all $p$ remains poor as shown in figure \ref{fig:gdcomparison}.

% the edges are already "mixed model"
We find in particular that the agreement as a function of $p$ is unsatisfactory. When considering the structure of the RG unit in figure \ref{fig:rgunitcell}, we note that the four corner nodes had been chosen to correspond to $t=0$ and $t=1$ in the opposing corners of the RG unit. So instead of simply ignoring these four corner nodes and concentrating on the five-node RG unit, it would be equally justified to include them into a now nine-node RG unit in which the four corner nodes are already part of the mixed model. This then suggests that one might introduce a normalized $p'= \frac{1}{2} \left( \frac{4}{9} + \frac{5 (2p)}{9} \right) = \frac{2}{9} + \frac{5 p}{9}$ with the $\frac{1}{2}$ in the initial expression chosen such that $2/9 \leq p'\leq 1/2$ also. In figure \ref{fig:gdcomparison}, we show the results for $\nu_{p'}$. 
Clearly, $\nu_{p'=2/9}$ is now in better agreement with the value of \citet{Klumper2019RandomGeometry}. But for other $p'$ values, large differences remain.

% small RG unit and large influence of the central node
Another source of systematic deviation from these results might be the small size of the RG unit and in particular, the privileged position of the central node, i.e. node three in figure \ref{fig:rgunitcell}. Effectively, it is the tunneling across this node which allows for self-interference in a figure-of-eight arrangement \cite{Galstyan1997LocalizationApproach}. In our implementation of the mixed model, we have thus far allowed this node to also "percolate" with $t=0$ and $1$. Excluding this possibility requires to rescale $p$ via $p^{*}= 
%\frac{1}{2} \left( \frac{4}{9} + \frac{4(2p)}{9} \right) =
\frac{2}{9} + \frac{4p}{9}$ such that ${2}/{9} \leq p^{*} \leq 4/9$. 
Our numerical results for $\nu$ both with and without this modification are displayed in table \ref{tab:gdresults}. The modification results in even larger values for $\nu$ than previously, as shown in figure \ref{fig:gdcomparison}. Clearly, we seem stuck in a situation where the RG approach cannot reproduce the results of the mixed model.
 %%%%%%%%%%%%%%%%%%%%%%%%%%%%%%%%%%%%%%%%%%%%%%%%%%%%%%%%%%%%%%%%%%%
% %%%%%%%%%%%%%%%%%%%%%%%%%%%%%%%%%%%%%%%%%%%%%%%%%%%%%%%%%%%%%%%%%%%

\renewcommand{\arraystretch}{1.5}
%%%%%%%%%%%%%%%%%%%%%%%%%%%%%%%%%%%%%%%%%%%%%%%%%%%%%%%%%%%%%%%%%%%
\begin{table}[t]
\centering
    \begin{tabular}{cccc}
        %\multicolumn{3}{c}{GD for RSRG unit as in Fig.\ \ref{fig:rgunitcell}} \\
        %\hline 
         $p$ & $p'$ & $p^{*}$ & $\nu$ \\
        \hline
        \hline
          0 & 0.222 & & $2.51\substack{+0.11\\-0.11}$  \\
         %\hline
          0.01 & 0.228 & & $2.58\substack{+0.03\\-0.03}$ \\
         %\hline
          0.025 & 0.236 & & $2.83\substack{+0.13\\-0.4}$ \\
         %\hline
          0.075 & 0.263 & & $4.1\substack{+0.7\\-0.4}$ \\
         %\hline
          0.1 & 0.278 & & $4.9\substack{+0.6\\-0.6}$ \\
         \hline
          0.1 & & 0.267 & $4.2\substack{+0.4\\-0.4}$ \\
         %\hline
         0.2 & & 0.311 & $11.9\substack{+1.2\\-1.4}$\\
         %\hline
          0.3 & & 0.356 & $25\substack{+11\\-9}$ \\
         \hline\hline
    \end{tabular}
    \caption[Results for $\nu$ including Geometric Disorder]{Values attained for the critical exponent $\nu$ when implementing the mixed disorder model at varying proportion levels $p$, $p'$ and $p^{*}$ in the RSRG. Data below the horizontal separator are generated excepting the central node from the mixing of models.}
    \label{tab:gdresults}
\end{table}
%\begin{table}[t]
%\centering
%    \begin{tabular}{cccc}
%        %\multicolumn{3}{c}{GD for RSRG unit as in Fig.\ \ref{fig:rgunitcell}} \\
%        %\hline 
%         $p$ & $p'$ & $p^{*}$ & $\nu$ \\
%        \hline
%        \hline
%          0 & 0.222 & & $2.51\substack{+0.11\\-0.11}$  \\
%         %\hline
%          0.01 & 0.225 & & $2.58\substack{+0.14\\-0.14}$ \\
%         %\hline
%          0.025 & 0.229 & & $2.8\substack{+0.2\\-0.4}$ \\
%         %\hline
%          0.075 & 0.243 & & $4.1\substack{+0.7\\-0.4}$ \\
%         %\hline
%          0.1 & 0.25 & & $4.9\substack{+0.6\\-0.6}$ \\
%         \hline
%          0.1 & & 0.244 & $4.2\substack{+0.4\\-0.4}$ \\
%         %\hline
%         0.2 & & 0.267 & $12\substack{+2\\-2}$\\
%         %\hline
%          0.3 & & 0.289 & $25\substack{+20\\-9}$ \\
%         \hline\hline
%    \end{tabular}
%    \caption[Results for $\nu$ including Geometric Disorder]{Values attained for the critical exponent $\nu$ when implementing the mixed disorder model at varying proportion levels $p$, $p'$ and $p^{*}$ in the RSRG. Data below the horizontal separator are generated excepting the central node from the mixing of models.}
%    \label{tab:gdresults}
%\end{table}
%%%%%%%%%%%%%%%%%%%%%%%%%%%%%%%%%%%%%%%%%%%%%%%%%%%%%%%%%%%%%%%%%%%

%%%%%%%%%%%%%%%%%%%%%%%%%%%%%%%%%%%%%%%%%%%%%%%%%%%%%%%%%%%%%%%%%%%
\section{Conclusion}
%%%%%%%%%%%%%%%%%%%%%%%%%%%%%%%%%%%%%%%%%%%%%%%%%%%%%%%%%%%%%%%%%%%
The localisation length critical exponent, $\nu$, of the quantum Hall transition can be estimated with the RSRG applied to the Chalker-Coddington model. 
Increasing the number of the statistical sample sizes 500 times, and using a more precise model for numerical data, has allowed us to get more stable and reliable estimates for $P(z)$. This in turn has enabled us to study the RSRG description of $\nu$ closer to the fixed point, being able to reduce $z_0$ by a factor of ten when compared to earlier RSRG approaches \cite{Cain2003a}. We find an increased estimate of $\nu\approx 2.51\substack{+0.11\\-0.11}$. 
This is in line with the general trend observed in the past $2$-$3$ decades when studying critical properties of second-order (quantum) phase transitions: substantially better statistics coupled with larger system sizes shift critical exponents towards $\sim 10\%$ larger values. This is true for the exponent of the QHE \cite{Slevin2009CriticalTransition,Puschmann2019IntegerLattice} as well as the 3D exponent of the Anderson metal-insulator transition \cite{Slevin1999a,Rodriguez2010,Rodriguez2011}.
A recent proposal for a deformed Wess-Zumino-Novikov-Witten type conformal field theoretic description of the QHE implies a resulting critical exponent $1/\nu \rightarrow {0}$ \cite{Zirnbauer2021MarginalTransition}, which would call for truly challenging computational resources or novel insight to allow a numerical validation.

For the mixed disorder model, proposed in reference \cite{Gruzberg2017GeometricallyTransitions}, our results are inconclusive. Clearly, the model has its charms and attractions when discussed in connection to experimental results for the QHE. However, our current, very local RSRG cell fails to capture the expected behaviour of $\nu$ and returns rather unphysical results. Previous attempts to improve upon this by enlarging the RG cell have been shown to fail due to the inherent difficulties associated with such an approach \cite{Assi2019ADisorder}.

%%%%%%%%%%%%%%%%%%%%%%%%%%%%%%%%%%%%%%%%%%%%%%%%%%%%%%%%%%%%%%%%%%%
\section{Acknowledgments}
%%%%%%%%%%%%%%%%%%%%%%%%%%%%%%%%%%%%%%%%%%%%%%%%%%%%%%%%%%%%%%%%%%%
We thank Warwick's Scientific Computing Research Technology Platform for further computing time and support. UK research data statement: Data accompanying this publication are available at \cite{ShawReal-spacePortal} while the code is at \cite{ShawDisQS/CCxD:Models}.

%%%%%%%%%%%%%%%%%%%%%%%%%%%%%%%%%%%%%%%%%%%%%%%%%%%%%%%%%%%%%%%%%%%
\printcredits

%%%%%%%%%%%%%%%%%%%%%%%%%%%%%%%%%%%%%%%%%%%%%%%%%%%%%%%%%%%%%%%%%%%
%% Loading bibliography style file
\bibliographystyle{model1-num-names}
%\bibliographystyle{cas-model2-names}

% Loading bibliography database
%\bibliography{references}
%\bibliography{referencesRAR}

%\vskip3pt

\end{document}